\documentclass{kapproc} % Computer Modern font calls

\def\solm{M$_{\odot}\,$}

\usepackage{procps} 

\usepackage[dvips]{graphicx}

\upperandlowercase

\kluwerbib % will produce this kind of bibliography entry:

\begin{document}

%------------ article title  ------------------->>

% If you use \\'s , please supply an alternate version of the title
% in square brackets, i.e., 
%\articletitle[Communism, Sparta, and Plato]
%{COMMUNISM, SPARTA,\\ and PLATO}

\vskip -2in
\articletitle[]{The Formation of the Hubble Sequence}
%% optional, to supply a shorter version of the title for the running head:
%%\chaptitlerunninghead{}

%%\author{Christopher J. Conselice}

%% multiple authors may be separated with \\
%% \author{Samuel Bostaph\\
%% and Gregor Kariotis}

%------ author/affiliation choices -------------->>

%% Single author

 \author{Christopher J. Conselice}
 \affil{California Institute of Technology}
 \email{cc@astro.caltech.edu}

%% Multiple authors, single affiliation

% \author{Samuel Bostaph}
% \author{George Lewis}
% \and   % <=== Type in \and before the last author so that `and' will
% \author{Cleon Jones}    % print between the last two authors
                           % in the table of contents.

% \affil{}

%------ prologue, abstract, keywords ----------->>
% optional prologue
%\prologue{<text>}{<author, year>}

% optional abstract
 \begin{abstract}
The history of galaxy formation via star formation 
and stellar mass assembly rates is now known with some certainty, yet
the
connection between high redshift and low redshift galaxy populations
is not yet clear.  By identifying and studying individual massive galaxies at 
high-redshifts, $z > 1.5$, we can possibly uncover the physical effects  
driving galaxy formation.  Using the structures of high-z galaxies,
as imaged with the Hubble Space Telescope, we argue that it is now possible to 
directly study the progenitors of ellipticals and disks.   We also 
briefly describe early results that suggest many massive galaxies are forming 
at $z > 2$ through major mergers.

\end{abstract}

% optional keywords
% \begin{keywords}
% Text, text...
% \end{keywords}

%------------ body of article ------------------->>

\subsection{Introduction}

Astronomers have come a long way in the last decade towards detecting and
studying galaxies out to redshifts $z \sim 6$.  However, the principle goal of 
observational galaxy formation studies, understanding how high 
redshift and low redshift populations are connected, and thus {\em how} 
galaxies form, remains largely unknown.   The traditional method of
studying galaxy formation is to try and answer {\em when} galaxies
formed, e.g., through star formation and stellar mass assembly histories and
by detecting various galaxy populations such as extremely red objects, 
Lyman-alpha emitters, and
Lyman-break galaxies. These galaxies, and their stellar populations, 
tell
us when galaxies formed, but they do not, and empirically cannot,
tell us {\rm how} they formed.

An approach to this problem is to use ancillary information
about galaxies not available from spectroscopy or integrated
photometry.  One method is to utilize structural
information from high-$z$ galaxies, imaged with HST, to determine when 
normal galaxies (hereafter defined as giant ellipticals and spirals) formed, 
as well as how.  
Techniques for doing this are now developed (Conselice 2003) and
early results demonstrate that we can identify and
trace the physical processes responsible for the formation of
galaxies at $z > 1.5$.  In this article we briefly describe these 
approaches for understanding the formation of the Hubble sequence, or normal
galaxies,  and give some preliminary results.

\subsection{Galaxies Near and Far}

Bright and massive galaxies in the nearby universe are disks
and ellipticals.  At higher redshifts, the brightest galaxies in rest-frame
UV selected samples nearly all have structural 
peculiarities (Conselice et al. 2003a).  When did normal disks and ellipticals
form?  Figure~1a demonstrates through
an I $< 27$ selected sample in the Hubble Deep Field North (HDF-N) that the
high redshift galaxy population is dominated by peculiars, while
at low-$z$ ellipticals and spirals are common.  
This transition is robust to effects of resolution,
noise, and morphological k-corrections (Conselice et al. 2004, in prep.)

A fundamental question to ask is whether these high redshift peculiars, often
called Lyman-break galaxies (LBGs), transform into disks and ellipticals.   
LBGs are likely the progenitors of massive nearby systems, based
on their clustering properties (Giavalisco et al. 1998).  However,  LBGs are 
observed to have low stellar masses, generally $<$ M$^{*}$ (e.g., Papovich 
et al. 2001).  If LBGs passively evolve, they will
not contain enough mass to become massive, $>$ M$^{*}$, galaxies at $z \sim 0$.
If we can determine the future evolution of
LBGs, and other high redshift galaxy populations, we can begin to
piece together the history of galaxy formation.

\subsection{The Galaxy Merger History}

There are a few major methods by which galaxies can form.  The first
is an early collapse when stars form in rapid bursts which
then passively evolve. Other methods are due to hierarchical
accretion of intergalactic material, or mergers with other galaxies.
The formation methods of galaxies are hard to trace, although based on the
star formation history of the universe it is unlikely
that most galaxies formed rapidly and early (e.g., Madau et al.
1998).  

One method that can be traced is the incidence
of major mergers.  Major mergers in the nearby universe create
distinct disturbed asymmetric morphologies that can be distinguished
from pure star forming systems through the use of indices that
measure asymmetries and the clumpiness of light distributions (Conselice
et al. 2000; Bershady et al. 2000; Conselice 2003).   The basic idea is that 
galaxies which are asymmetric, without a corresponding high
degree of light clumpiness, are likely to be involved in major
mergers (Conselice 2003).  How well does the asymmetry index (A) 
identify known mergers? Figure 1b shows the deviation in sigma units
between asymmetry (A) and clumpiness (S) values for nearby
normal galaxies and major mergers. While normal galaxies have
mostly low $\sigma$ deviations from the A-S correlation, major mergers span a
much larger range and include the most asymmetric objects. 

We can use asymmetry values of HDF-N galaxies to measure the 
evolution of implied major merger fractions out to $z \sim 3$. We avoid 
morphological K-corrections by using the rest-frame B-band
asymmetries of galaxies in the HDF-N (Conselice et al. 2003a).
We define major mergers as galaxies with rest-frame
B-band asymmetry larger than $A_{\rm merger} = 0.35$. By taking the ratio of 
galaxies with asymmetries 
$A > A_{\rm merger}$ to the total number of galaxies in a given parameter 
range, implied merger fractions out to $z \sim 3$ can be computed as a 
function of absolute magnitude (M$_{\rm B}$), stellar mass (M$_{*}$), and 
redshift (z). This allows us 
to determine how and when galaxies formed as a function of stellar mass
and time.

\begin{figure}
%\hspace{2cm}
\vskip -2in
\includegraphics[width=5in]{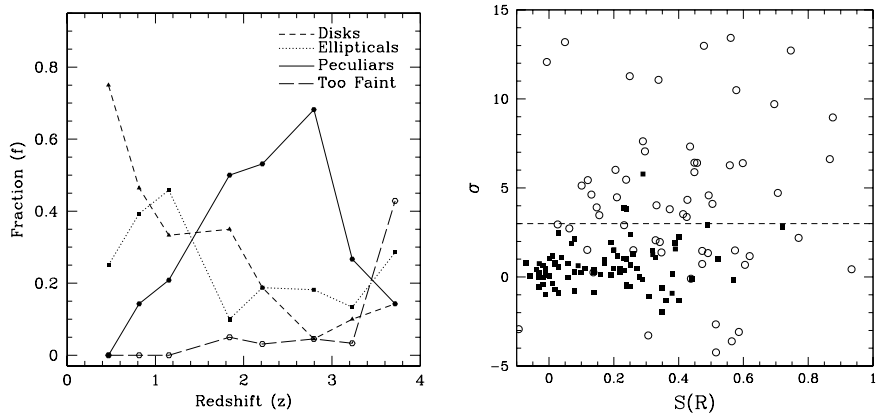}
\vskip -2.5in
\vskip.2in
\caption{Left panel (a) - the morphological distribution as a function of
type in the HDF, normalized to I$< 27$ counts.  Right panel (b) - the sigma
deviation from the asymmetry-clumpiness correlation, showing that major mergers
(open circles) deviate by more than $3 \sigma$ from this relationship, while 
normal galaxies (solid boxes) generally do not.}
\vskip -0.3in
\end{figure}

Major merger fractions for galaxies brighter than M$_{\rm B} = -21$ and $-19$ 
in the HDF-N are plotted in Figure~2a, along with 
semi-analytical model predictions of the same quantities.  From this it 
appears that a large fraction of the brightest galaxies are undergoing 
mergers, while fainter 
systems generally do not. The fraction of bright galaxies undergoing major
mergers drops quickly at lower redshifts, while the fainter systems
have merger fractions that remain largely constant with redshift (Conselice
et al. 2003a).
Semi-analytic model predictions of the same quantities, based on Durham 
group simulations, 
do a relatively good job of predicting merger fractions at high redshift, but 
over predict the degree of major
merging at lower redshifts for the most luminous and massive galaxies 
(Figure~2a).

\subsection{Merger and Stellar Mass Accretion Rates}

Using further information we can investigate the merger and stellar mass 
accretion rates of galaxies from $z = 0$ out to $z \sim 3$, or to when the
universe was only $\sim 2$ Gyr old.  By assuming
a merger time scale of 1 Gyr, and that each 
merger consists of two galaxies of equal mass, we can measure
the rate of merging, and the stellar mass accretion rate due to 
major mergers.  We can also determine how this accretion
history varies with initial stellar mass (Figure 2b; Conselice et al.
2003a,b).  Based on these empirical measurements, the amount of stellar mass
added to a M $\sim 10^{10}$ \solm LBGs due to star formation and merging is 
enough to create a massive $>$ M$^{*}$ galaxy at $z \sim 0$ 
(Conselice 2004, in prep).

\begin{figure}
%\hspace{2cm}
\vskip -2.4in
\includegraphics[width=5in]{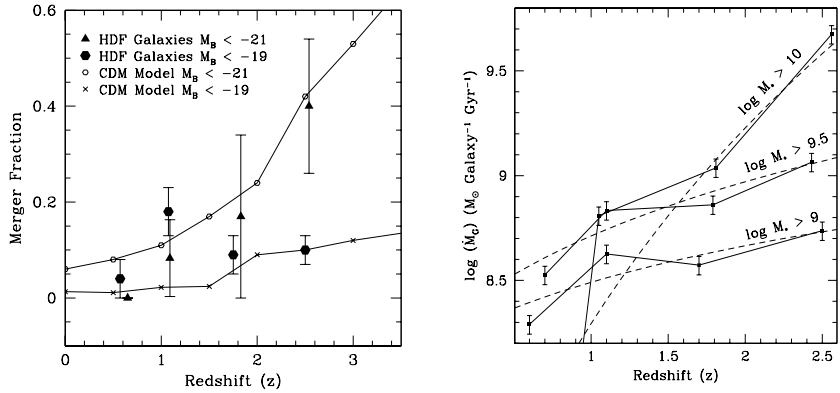}
\vskip -2.5in
\vskip.24in
\caption{Left panel (a) - major merger fractions to $z \sim 3$ at
magnitude limits M$_{\rm B} = -21$ and $-19$. Semi-analytical model
predictions are also shown.  Right panel (b) - stellar mass accretion
history from major mergers as a function of initial mass 
(see Conselice et al. 2003a).}
\vskip -0.2in
\end{figure}

Disk galaxies likely cannot form through these merger processes as they 
would not easily survive mergers at high redshift.  
These systems are therefore likely forming
at about the same time they appear morphologically, at 
$z \sim 1.5$ (Figure~1).   These galaxies have now possibly
been identified at $z > 1.5$ by their low light concentrations 
in GOODS ACS images (Conselice et al. 2003c).
These luminous diffuse objects (LDOs) are common at redshifts
$1 < z < 2$ and have co-moving volumes similar to nearby massive
disks.   Follow up on these systems is now in progress.


\begin{chapthebibliography}{<widest bib entry>}
\bibitem[]{} Bershady, M.A., Jangren, A., \& Conselice, C.J. 2000, AJ, 126, 1183
\bibitem[]{} Conselice, C.J., et al. 2000, ApJ, 529, 886 
\bibitem[]{} Conselice, C.J. 2003, ApJS, 147, 1 
\bibitem[]{} Conselice, C.J., et al. 2003a, AJ, 126, 1183 
\bibitem[]{} Conselice, C.J., Chapman, S.C., \& Windhorst, R.A. 2003b, ApJ, 596, 5L 
\bibitem[]{} Conselice, C.J., et al. 2003c, ApJ in press, astro-ph/0309039 
\bibitem[]{} Giavalisco, M., et al. 1998, ApJ, 503, 543 
\bibitem[]{} Madau, P., Pozzetti, L., \& Dickinson, M. 1998, ApJ, 498, 106 
\bibitem[]{} Papovich, C., Dickinson, M., \& Ferguson, H.C. 2001, ApJ, 559, 620 
\end{chapthebibliography}
\end{document}